\documentstyle[11pt,epsfig]{article}
\begin{document}   
\bibliographystyle{plain}

\def\RR{\rm \hbox{I\kern-.2em\hbox{R}}}
\def\NN{\rm \hbox{I\kern-.2em\hbox{N}}}
\def\ZZ{\rm {{\rm Z}\kern-.28em{\rm Z}}}
\def\CC{\rm \hbox{C\kern -.5em {\raise .32ex \hbox{$\scriptscriptstyle
|$}}\kern
-.22em{\raise .6ex \hbox{$\scriptscriptstyle |$}}\kern .4em}}
\def\L{\pounds}
\def\<{\langle}
\def\>{\rangle}
\def\nl{\newline}
\def\vp{\varphi}
\def\e{\varepsilon}
\def\s{{\cal S}}
\def \vo {{\mathbf{\omega}}}
\def \ve {{\mathbf{\epsilon}}}
\def \mL {{\mathbf{\Lambda}}}
\def \mLt {{\mathbf{\Lambda}^t}}
\def \mLi {{\mathbf{\Lambda}^{-1}}}
\def \mS {{\mathbf{\Sigma}}}
\def \mSt {{\mathbf{\Sigma}^t}}
\def \mSi {{\mathbf{\Sigma}^{-1}}}
\def \Cov {{\mbox{Cov}}}
\def \Var {{\mbox{Var}}}

\newtheorem{Tm}{Theorem}[subsection]
\newtheorem{Df}[Tm]{Definition}
\newtheorem{Lm}[Tm]{Lemma}
\newtheorem{Cr}[Tm]{Corollary}
\newcommand{\Rq}{\addtocounter{Tm}{1}{\bf Remark \theTm } \hspace{0.15cm}}
\newcommand{\Ex}{\addtocounter{Tm}{1}{\bf Example \theTm } \hspace{0.15cm}}
\newenvironment{Pf}{\begin{quote}{\bf Proof} \hspace{0.03cm}}{{$\Box $}
\newcommand{\ss}{\scriptscriptstyle}
\newcommand{\s}{\scriptstyle}
\end{quote}}

\title{Modelling fluctuations of financial time series: from cascade
process to stochastic volatility model}

\author{J.F. Muzy$^{1}$, J. Delour$^{1}$ and E. Bacry$^{2}$ \\
$^1$ Centre de Recherche Paul Pascal, Avenue Schweitzer 33600 Pessac,
France \\
$^2$ Centre de Math\'ematiques appliqu\'ees, Ecole Polytechnique,  91128
Palaiseau Cedex, France}

\date{\today}

\maketitle

\abstract{
In this paper, we provide a simple, ``generic'' interpretation
of multifractal scaling laws and
multiplicative cascade process paradigms in terms
of volatility correlations. We show that in this context
1/f power spectra, as recently observed in Ref.~\cite{aBon99},
naturally emerge. We then propose
a simple solvable ``stochastic volatility'' model for return fluctuations.
This model is able to reproduce most of recent empirical findings concerning
financial time series: no correlation between price variations,
long-range volatility correlations and multifractal statistics.
Moreover, its extension to a multivariate context, in order to
model portfolio behavior, is very natural.
Comparisons to real data and other models proposed elsewhere
are provided.
}

\section{Introduction}
\label{intro}
As shown by most recent empirical studies on
huge amount of data, the market price changes are
characterized by several ``universal''
features \cite{bBou99,bMan00}: price increments are
not correlated, volatilities are strongly (power-law)
correlated and price increment probability density function (pdf)
shapes depend on
the time scale. From quasi Gaussian at rather
large time scales, these pdf are characterized by fat tails
at fine scales. Many authors
in the recently emerged field of ``econophysics''
\cite{bBou99,bMan00,aFar99}
as well as
in classical empirical finance, aim at proposing simple, discrete
or continuous time models
that are able to account for these observations.
Among all the proposed models, one can distinguish several
streams, from the simplest Brownian process, that
constitutes the main tool used by practitians, to
the class of ``heteroskedastic'' nonlinear processes
as proposed in Refs. \cite{aEng82,aBol92}.
To account for the letpokurtic nature of the
small scale pdf,
Mandelbrot \cite{aMan63} and Fama \cite{aFam65}
proposed the Levy stable paradigm that has been recently improved
in the ``truncated Levy'' version \cite{aMan95,bBou99,bMan00}.
More recently, an interesting comparison
between market price variations and the fluctuations of the
fluid velocity field in fully developed turbulence has been
suggested \cite{aGha96}. Besides the real pertinence of such
an analogy that has been
widely commented \cite{aMan96,pArn96,bMan00}, this work
opens the door to another important paradigm to model financial
time series, namely {\em multifractal processes}.
The multifractal processes\footnote{people sometimes refer
  to ``multi-affine'' processes or processes that display
``multi-scaling''},
and the deeply connected mathematics of large deviations and
multiplicative cascades, are well known to be useful to
describe the intermittent nature of fully developed turbulence
\cite{bFri95}.
Recent empirical findings \cite{aArn98,pFis97,pBra99,pSch99}
suggest that in finance, this framework is also likely to be pertinent
as far as the time scale dependence of
the statistical properties of price variations is concerned.

Our purpose in this paper is twofold. First, we make
a brief review of multifractals in order to specify what is a
multifractal process. We try to provide several complementary
points of view and to understand what are the main ingredients for
``multi-scaling''.
We also comment about the criticisms raised by several
authors about multifractality in finance.
Our second goal is to propose a simple
multifractal ``stochastic volatility'' model that
captures very well all the above mentionned features of financial
fluctuations.
This model, that has been originally introduced in Ref. \cite{prl},
is compared to real data and some models proposed elsewhere.
We discuss its possible multivariate extension
in order to use it in management applications.
The paper is organized as follows.
The review on multifractal processes is made in section 2.
We introduce notations, the related
notions of multi-scaling, scale-invariance,
cascade process and self-similarity kernel.
We illustrate our purpose using empirical
estimates for some high frequency financial data.
In section 3 we review some findings of Ref. \cite{aArn98}
concerning the magnitude correlations for cascade
models and suggest a link with $1/f$ processes
as recently observed in financial time series.
In section 4 we introduce the multifractal random walk
defined in Ref. \cite{prl} as a stochastic volatility
model. We discuss its main properties
and propose a natural multivariate generalization. Our discussion
is illustrated by numerical simulations.
In section 5 we propose estimators for the few parameters of
our model and compute them for some intraday and daily
time series.
In section 6 we discuss some related works about
multifractality in finance.
Conclusions and some prospects are reported in section 7.

\section{Multifractal processes and cascade models}
\label{sec1}
In this section we briefly discuss the related notions
of multifractality and multiplicative
cascade. Most of the ideas and concepts that we recall
below have been introduced in the field
of fully developed turbulence where people aim
at accounting for the so-called ``intermittency phenomenon''
(for a review of this subject see e.g., \cite{bFri95}).

\subsection{Multifractality of financial time series}
Let us consider the variations of
a stochastic process $X(t)$ at a time scale $l$.
For that purpose, one can consider the increments
of the process, $\delta_l X(t) = X(t+l) - X(t)$ or
more generally its wavelet transform \cite{WT1,WT2,WT3}
$$
T(t,l) = l^{-1} \! \int \psi\left(\frac{t'-t}{l}\right) X(t') dt'
$$
where $\psi(t)$ is the so-called
analyzing wavelet, i.e, a function
well localized in both Fourier and direct spaces\footnote{One nice property
of wavelet transform is that it can be inverted, i.e., one can recover the
original signal from its wavelet coefficients. Another interesting
feature is that there exist orthonormal wavelet bases. Such bases
are very useful for signal synthesis and modelling, as it is
illustrated for cascade processes in Ref. \cite{jmp}}.
Let us denote $M(q,l)$ the order $q$ absolute moment
of $\delta_l B(t)$ or $T(t,l)$,
(in this paper $E(.)$ will be used for the mathematical expectation
and we will always suppose that the considered processes has
stationary increments)
\begin{equation}
 M(q,l) = E(|\delta_l X(t)|^q) \; .
\end{equation}
\begin{figure}
  \begin{center}
    \includegraphics{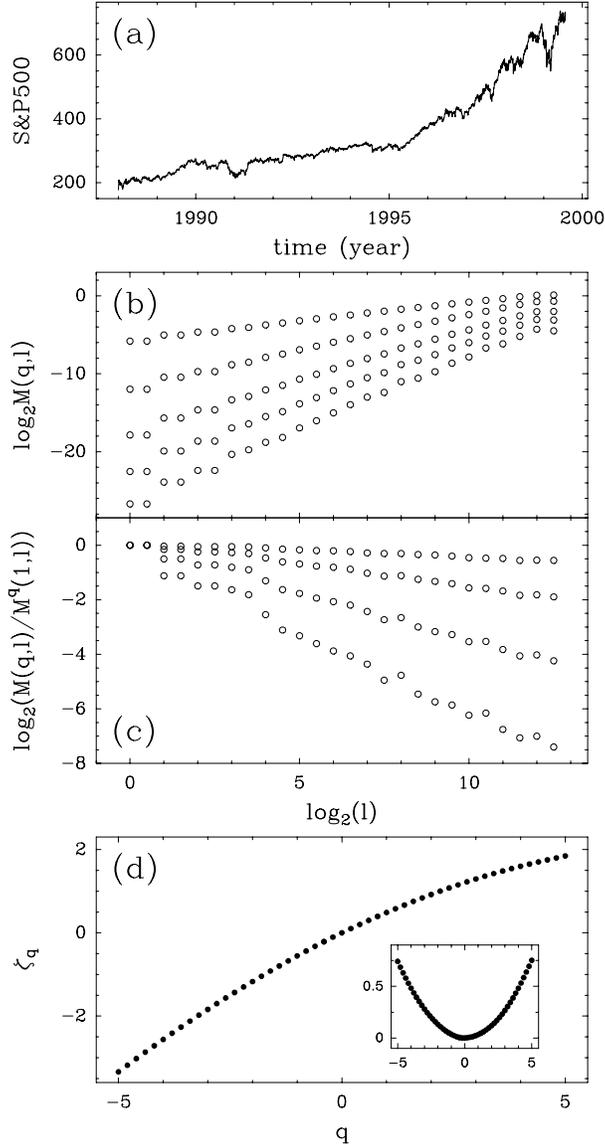}
    \caption{
      {\bf Multifractal Analysis of the intraday
        future S\&P500 index over the period 1988-1999}.
      (a) Plot of the original index time-series. The analyzed
       time-series is the detrended and de-seasonalized
       logarithm of this series.
      (b) Log-log plots of $M(q,l)$ versus $l$ for $q=1,2,3,4,5$.
       The time scales $l$ range from 10 minutes to 1 year.
      (c) $\log_2(M(q,l)/M(1,l)^q)$ for $q=2,3,4,5$.
      Such plots should be horizontal for a process that is not
multifractal.
      (d) $\zeta_q$ spectrum for the S\&P 500 fluctuations.
       The plot in the inset is the parabolic nonlinear part
       of $\zeta_q$.
      }  
    \label{fig_1}
  \end{center}
\end{figure}
We will say that the process is {\em scale-invariant},
if the scale behavior of the absolute moment $M(q,l)$
is a power law. Let us call $\zeta_q$ the exponent of this
power law, i.e., 
\begin{equation}
  M(q,l)  \sim C_q l^{\zeta_q},
\label{scaling}
\end{equation}
where $C_q$ is a prefactor that will be interpreted below.
The process is called {\em monofractal} if
$\zeta_q$ is a linear fonction of $q$ and {\em multifractal}
if $\zeta_q$ is nonlinear. Note that,
from the concavity of the moments of a random variable, it is easy
to show that $\zeta_q$, as defined from the scaling behavior (\ref{scaling})
in the limit $l \rightarrow 0^+$, is necessarily a convex fonction of $q$.
The same argument leads to the conclusion that such scaling behavior
with a nonlinear $\zeta_q$ cannot hold for all scales $l$. Thus, for a
multifractal process there exists at least one characteristic time $T$
(hereafter referred to as the {\em integral time}) above
which the behavior (\ref{scaling}) is no longer valid. Multifractality has
been introduced in the context
of fully developed turbulence in order
to describe the spatial fluctuations of the fluid velocity
at very high Reynolds number \cite{bFri95}.
As suggested by recent studies \cite{aGha96,aArn98,pFis97,pBra99,pSch99},
multifractality is likely to be a pertinent concept to account
for the prices fluctuations in financial time-series.
This is illustrated in Fig. \ref{fig_1} where
the $\zeta_q$ function is estimated for the future S\&P500 index
over the period 1988-1999. The original intraday time-series
has been sampled at a $10$ mn rate (Fig. \ref{fig_1}(a))
in order to obtain equi-sampled data.
We consider the associated continuously compounded return time-series,
(i.e., the logarithm of the index value) that has been detrended
and de-seasonalized\footnote{The amplitude of the return variations
in each intraday period is normalized according to the
estimated U-shaped intraday r.m.s.}.
The $\zeta_q$ spectrum in Fig. \ref{fig_1}(c) is
obtained using linear regression fit
of ``log-log'' representations of the behavior of
the $q$-th order moment versus the time scale as illustrated in
Fig. \ref{fig_1}(b). In this figure, the scales
span an interval from 10 minutes to approximately 1 year.
Moment estimates at larger time scales are very poor
because of the finite size of the overall record.
From the linear behavior of such
curves, one clearly sees that the scale-invariance hypothesis is
satisfied over around 3 decades.
In Fig. \ref{fig_1}(b) we have plotted
$\log_2 \frac{M(q,l)}{M(1,l)^q}$ versus $\log_2(l)$.
The fact that such plots are not constant reflects the
nonlinearity the $\zeta_q$ spectrum. The future S\&P500
can thus be considered,
at least at this description level, as a multifractal
signal.
Let us notice that we have computed, in Fig. \ref{fig_1}(d),
the $\zeta_q$ values for $q$-th order moments that include
negative values of $q$. This can be achieved using
a wavelet based technique that has been introduced in
Refs. \cite{wtmm1,jsp,bifc,wtmm2}.
This spectrum turns out to be well fitted by a parabolic
shape $\zeta_q = 0.53 q - 0.015 q^2$. The non linear parabolic
component of $\zeta_q$ has been plotted in the inset
of Fig. \ref{fig_1}(d).

\subsection{Multifractal processes, self-similar processes
and multiplicative cascades.}
Multifractality (in the sense defined above)
is a notion that is often related to
an underlying multiplicative cascading process.
In the context of deterministic functions the situation
is rather clear since the analyticity of the $\zeta_q$
spectrum  is deeply connected to the self-similarity properties
of the function \cite{jsp,bifc,wtmm2,jaff}.
Roughly speaking, a function is {\em self-similar}
if it can be written as a multiplicative cascade in an
appropriate space-scale (or time-scale) representation
\cite{bifc,jsp,jaff}.
In that context, the so-called {\em multifractal
formalism} is valid, i.e., one can relate the $\zeta_q$ spectrum
to the $D(h)$ singularity spectrum that provides information
about the statistical distribution of singularity (H\"older) exponents $h$.
The things are somehow more complex for stochastic processes.
One of the goals of this paper is to provide some
simple elements about this subject.

In the mathematical literature, a process $X(t)$
is called self-similar of exponent $H$
if $\forall \lambda > 0 $, $ \lambda^{-H} X(\lambda t)$ is the
same process as  $X(t)$.
According to this definition, the Brownian
motion is self-similar with an exponent $H=1/2$.
This definition is however too restrictive for our purpose since
it excludes multifractal processes. Indeed, let us consider
$P_l(\delta X)$
the probability density function (pdf) of
$\delta_l X(t)$\footnote{Note that from stationarity of
the increments, the law of $\delta_l X(t)$ is
the same as the law of $X(l)$ if one assumes that $X(0)=0$.}.
If $X(t)$ is self-similar with an exponent $H$,
then it is easy to prove that
\begin{equation}
        P_l(\delta X) = \lambda^H P_{\lambda l} (\lambda^H \delta X) \; .
\label{ssm}
\end{equation}
Then, the moments at scale $l$ and $L = \lambda l$ are related by
\begin{equation}
         M(q,l) = C_q \left( \frac{l}{L} \right)^{qH} ,
\end{equation}
with $C_q = M(q,L)$. Thus one has a ``monofractal''
process with $\zeta_q = qH$.
In order to account for multifractality,
one has to generalize this classical definition
of self-similarity. This can be done
by introducing a weaker notion, as originally proposed
in the field of fully developed turbulence by B. Castaing
and co-authors \cite{castaing}.
According to Castaing's definition of self-similarity,
a process is self-similar if the increment pdf's at scales
$l$ and $L = \lambda l$ ($\lambda >1$) are related by
the relationship \cite{castaing,aArn97}:
\begin{equation}
    P_l(\delta X) = \int G_{l,L}(u) e^{-u} P_{L}(e^{-u} \delta X) du \; ,
\label{castaing}
\end{equation}
where the {\em self-similarity kernel} $G_{l,L}$ depends only on $l/L$. Let
us note that 
this definition generalizes Eq. (\ref{ssm}) that
corresponds to the ``trivial'' case $G_{l,L}(u) = \delta(u-H\ln(l/L))$.
This equation basically states that the pdf $P_l$ can be obtained through a
``geometrical convolution'' between the kernel $G_{l,L}$ and the pdf $P_L$.
A simple argument shows that the logarithm of the Fourier transform of the
kernel $G_{l,L}$
can be written as $F_{l,L}(k) = \ln \hat{G}_{l,L}(k) = F(k) \ln(l/L)$
\footnote{It essentially
results from the fact that $G_{l,L}$ depends only on $l/L$ and satisfies the
semi-group
composition law $G_{l_1,l_3} = G_{l_1,l_2} \ast G_{l_2,l_3}$
where $l_1 \leq l_2 \leq l_3$ and $\ast$ is the convolution
product \cite{castaing,aArn97}.}.
Thus, from Eq.~(\ref{castaing}), one can
easily show that the $q$ order absolute
moments at scales $l$ and $L$ are related by:
\begin{equation}
  \label{kernelscaling}
         M(q,l) = \hat{G}_{l,L}(-iq) M(q,L) =
M(q,L)\left(\frac{l}{L}\right)^{F(-iq)}  \; ,
\end{equation}
and then $C_q= M(q,L)$ and $\zeta_q = F(-iq)$.
A nonlinear $\zeta_q$ spectrum
implies that $F$ is nonlinear and thus that $G$ is different from
a Dirac delta function\footnote{Note that from the above mentionned
  semi-group property, the Levy theorem \cite{feller} implies that
  $G$ is necessarily an infinitely divisible law}.
For example, the simplest non linear
case is the so-called log-normal model that
corresponds to a parabolic $\zeta_q$ function and thus to
a function $G$ that is Gaussian.

The equation (\ref{castaing}) can be interpreted
as follows: the pdf at scale $l$, $P_l$
is written as a weighted
superposition of the rescaled versions of the pdf at scale $L$,
$P_L$, the self-similarity kernel
$G_{l,L}$ being the associated distribution of weights. In the case of a
monofractal
process as described by Eq. (\ref{ssm}), a
single value of
$u$ is sufficient in the equation (\ref{castaing}) since
$P_l$ and $P_L$ have the same shape and differ only by the scale
factor $e^{-u} = (l/L)^H= \lambda^H$.
This explains the Dirac function for the kernel $G$.
This situation can be easily generalized
by considering other shapes for the kernel $G_{l,L}$.
In that case, the
shapes of the pdf $P_l$ across scales are no longer the same: when
going to small scales, fat tails emerge and the pdf become
strongly leptokurtic
(see Refs.~\cite{aGha96,castaing} or Fig.~\ref{fig_4}).

Let us now make the link with multiplicative cascades. This can
be easily done if one consider discrete scales $l_n = 2^{-n} L$.
Let us suppose that the local variation of the process
$\delta_{l_n} X$ at scale $l_n$ is obtained
from the variation at scale $L$ as
\begin{equation}
  \label{casc}
     \delta_{l_n}X(t) = \left(\prod_{i=1}^{n} W_i\right) \delta_{L}X(t)
\end{equation} 
where $W_i$ are i.i.d. random positive factors.
This is the cascade paradigm. Realizations of such processes
can be constructed
using orthonormal wavelet bases as discussed
in Ref. \cite{jmp}. If one defines the {\em magnitude} $\omega(t,l)$
at time $t$ and scale $l$ as the logarithm of ``local
volatility'' \cite{aArn98}:
\begin{equation}
 \label{defomega}
     \omega(t,l) = \frac{1}{2} \ln(|\delta_{l} X(t)|^2),
\end{equation}
then the previous cascade equation becomes a simple random
walk equation, at fixed time $t$, versus the logarithm of scales:
$$
 \omega(t,l_{n+1}) = \omega(t,l_{n}) + \ln(W_{n+1}) \; .
$$
If the noise $\ln W_i$ is normal $N(\mu,\lambda^2)$,
the pdf of $\omega$, $P_l(\omega)$,
thus satisfies a simple diffusion
equation with a Gaussian kernel:
\begin{equation}
    P_{l_n}(\omega) = \left(N(\mu,\lambda^2)^{\ast n} \ast p_{L} \right)
(\omega)
\end{equation}
where $\ast$ is the convolution product. Going back to the
original variable $\delta X$, the previous equation
corresponds exactly
to Castaing's formulation of self-similarity (\ref{castaing})
with the log-normal propagator:
$$
 G_{l_n,L} = N(\mu,\lambda^2)^{\ast n} = N(n\mu,n\lambda^2) \; .
$$
Conversely, let us consider a process that
satisfies Castaing's equation with a normal kernel $G$.
This means that one can write,
\begin{equation}
 \label{claw}
   \delta_l X(t) \equiv W  \delta_{2l} X(t)
\end{equation}
where $\equiv$ means the equality in law of the two random variables
and $W$ is a log-normal random variable which mean $\mu$
and variance $\lambda^2$ do not depend on $l$.
By iterating this equation $n$ times, one thus recover,
at least heuristically, the cascade equation (\ref{casc}).
Thus, the cascade picture across scales, constitutes a kind of
paradigm of non-trivial self-similar processes.
As explained in Ref. \cite{prl}, the problem with such processes
is that they involve representations (e.g., orthonormal wavelet bases)
that are constructed on a discrete set of scales (e.g.,
dyadic scales $l_n = 2^{-n}$) and in turn
cannot be invariant under continuous scale dilations.

\section{Magnitude correlations and 1/f spectra}
\begin{figure}
  \begin{center}
    \includegraphics{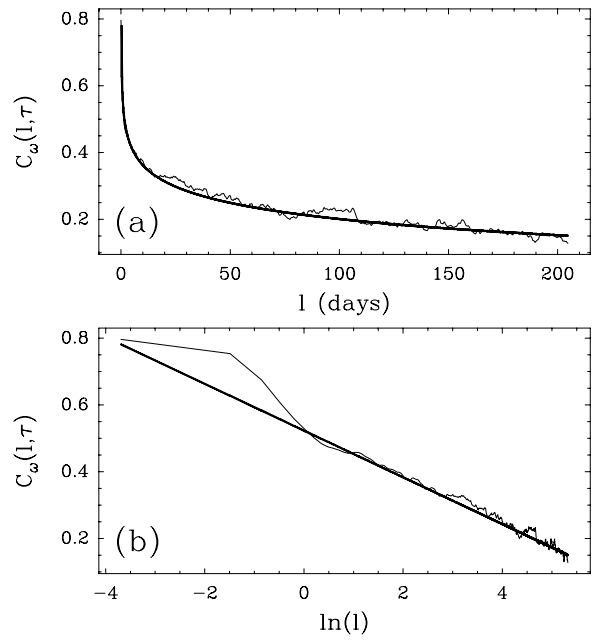}
    \caption{
      {\bf Magnitude correlation function of the
       S\&P 500 future}.
      (a) $C_{\omega}(l,\tau)$ versus $l$ for $\tau = 10$ min.
      The solid line represents a fit according to the cascade model
      logarithmic expression.
      (b) $C_{\omega}(l,\tau)$ versus $\ln(l)$. The cascade model
      predicts a linear behavior that crosses the y-axis at
      $\ln(l)=\ln(T)$. The small scale cross-over is due to
      the smoothing window used to estimate the local
      magnitude $\omega(t,\tau)$.
     }  
    \label{fig_2}
  \end{center}
\end{figure}

We have seen in the previous section that multifractality can be
interpreted as a diffusion of the magnitude of the variations
of the return from large time scales to small time scales.
In the financial framework, magnitudes at all scales
are nothing but a logarithmic representation of local volatilities.
In this section we would like to address the problem of
volatility correlations. The ``heteroskedastic'' nature
of financial time-series is now a well established empirical
fact. Volatility possesses long-range positive correlations:
periods of strong activity alternate with quiet periods.
A lot of models have been proposed to account for
this phenomenon from the famous GARCH models to various stochastic
volatility models. Let us proceed with the multifractal and cascade
picture and study what kind of correlations are associated to these
models. This problem has already
been considered by Arneodo, Muzy and Sornette
in Ref. \cite{aArn98} (see also Refs \cite{jmp,aArn98b}).
These authors have shown that a log-normal
cascade model on the dyadic tree associated to the orthonormal wavelet
representation leads naturally to magnitude correlation functions
$C_{\omega}(l,\tau) = \Cov(\omega(t,\tau),\omega(t+l,\tau))$
that behave as $-\lambda^2 \ln(l/T)$ for $T > l > \tau$.
This behavior has been shown to provide good fits of the empirical
estimates of the correlation functions from real data \cite{aArn98}.
In Fig. \ref{fig_2} is reported the magnitude correlation
function $C_{\omega}(l,\tau)$ (we choose $\tau = 10$ min)
of the S\&P500 time series studied in Fig. \ref{fig_1}.
One can see that, when plotted versus the logarithm of the time
lag, $\ln(l)$, the correlation function decreases
linearly with a slope $\lambda^2 \simeq 0.025$. The intercept of
such straight line
provides an estimator of the integral time $T$ that
is, in our case, approximately $T \simeq 3$ years (note that
because we get an estimate of $\ln(T)$ the error on the value of
$T$ is very large). We have checked that those results are
stable when changing the reference time $\tau$ for return calculation.
As it will be illustrated in section 5, for various financial time series,
the ``cascade ansatz'' is very pertinent
to describe the volatility correlations. Let us notice that
the very slow (logarithmic) decrease of the correlation functions for
time lags below the integral time $T$, is reminiscent of the ultrametric
nature of the tree naturally associated to the time-scale (or
time-frequency)
representation \cite{aArn98,jmp,aArn98b}.
Moreover, let us remark that,
as far as power spectrum is concerned, Gaussian processes
with such correlation functions can be seen as ``$1/f$'' processes.
Indeed, if the correlation function is given by the above expression,
the power spectrum can be shown to reduce to
\begin{equation}
   S(f) = 2\lambda^2 f^{-1} \int_0^{Tf} x^{-1} \sin(x) dx \; .
\end{equation}
In the high frequency limit $f \rightarrow +\infty$, we then
have $S(f) \sim \lambda^2 \pi f^{-1}$.
Another intuitive way to understand this property, comes from the fact that
the logarithmic decay of the correlation function
can be understood
as the limit $H \rightarrow 0$
in the power-law correlation function $k^{-2H}$
of a fractional gaussian noise of exponent $H$.
This property will be explicitely
used in the discussion of section 6.3.
Let us finally remark that
$1/f$ spectra have been observed
in a wide range of applications \cite{1fp}.
Recently, Bonanno {\em et al.} \cite{aBon99} suggested
the possible pertinence of such processes to account for the
fluctuations of the number of trades of different stocks.

\section{A simple solvable multifractal model}
As emphasized previously, multiplicative cascade models represent the
paradigm
of multifractal processes in that they contain the main
ingredient leading to multifractality, i.e, the scale evolution of
the magnitudes, from coarse to fine scales, is a random walk.
Besides the problems of continuous scale invariance and
stationarity of standard hierarchical constructions of
such processes, they cannot be formulated
using a stochastic evolution equation
as one would expect for a model for financial time series.
In this section we propose a ``stochastic volatility'' model
that has been introduced in Ref. \cite{prl},
that does not possess any of these drawbacks: it has  stationary
increments, it has log-normal multifractal properties and is invariant
under continuous dilations. The key idea underlying this
model is that the stochastic volatility possesses, as
for cascading processes, a ``$1/f$'' spectrum, or, more precisely,
a correlation function with a logarithmic behavior.

\subsection{The multifractal random walk}
Let us briefly recall the construction of the
{\em multifractal random walk} (MRW) proposed in \cite{prl}.
A discretized version of the model $X_{\Delta t}$
(using a time discretization step $\Delta t$)
is built by adding up $t/\Delta t$ random variables :
$$
X_{\Delta t} (t) = \sum_{k=1}^{t/\Delta t} \delta X_{\Delta
t}[k],
$$
where the process $\{\delta X_{\Delta t}[k]\}_k$ is a noise
whose variance is stochastic, i.e.,
\begin{equation}
  \label{model1}
\delta X_{\Delta t}[k] = \epsilon_{\Delta t}[k] e^{\omega_{\Delta t}[k]} \;
,
\end{equation}
where $\omega_{\Delta t}[k]$ is the logarithm of the stochastic
variance.
More specifically, we will choose $\epsilon_{\Delta t}$ to be a
gaussian white noise independent of $\omega$ and
of variance $\sigma^2 \Delta t$.
The choice for the process $\omega_{\Delta t}$ introduced
in \cite{prl}, is dictated by the cascade picture.
It corresponds to a gaussian stationary process
whose covariance can be written
$$\Cov(\omega_{\Delta t}[k],\omega_{\Delta t}[l]) =
\lambda^2 \ln \rho_{\Delta t} [|k-l|]$$
where $\rho_{\Delta t}$ is
chosen in order to mimic the correlation structure
observed in cascade models with an integral time $T$:
$$
\rho_{\Delta t} [k] =
\left\{
\begin{array}{ll}
\frac{T}{(|k|+1)\Delta
t} & \mbox{for}~|k|\le T/\Delta t -1
\\
1 & \mbox{otherwise}
\end{array}
\right.
$$
Hereafter, we will refer to the process $\omega(t)$ as the
``magnitude process''. In order the variance of $X_{\Delta t} (t)$
to converge when  $\Delta t\rightarrow 0$,
one must choose the mean of the process $\omega_{\Delta t}$
such that \cite{prl}
$$
E\left(\omega_{\Delta t}[k]\right) = -\Var\left(\omega_{\Delta t}[k]\right)
= - \lambda^2\ln(T/\Delta t),
$$
for which we find $\Var(X_{\Delta t}(t)) = \sigma^2t$.
Let us review the multifractal properties of MRW.

\subsection{$\zeta_q$ spectrum: computation of the moments}
The $q$th-order moment of the increments of the MRW can
be computed. Since, by construction, the increments of the model
are stationary, the law of $X_{\Delta t}(t+l)-X_{\Delta t}(t)$
does not depend on $t$ and is the same
law as $X_{\Delta t}(l)$.
In Ref. \cite{prl}, it is proven that the moments of
$X(l) \equiv X_{\Delta t \rightarrow 0^+}(l)$
can be expressed as
\begin{equation}
E(X(l)^{2p}) = \frac{\sigma^{2p}(2p)!}{2^p p!} \int_{0}^{l}du_1
...\int_{0}^{l}du_p
\prod_{i<j} \rho(u_i-u_j)^{4\lambda^2},
\end{equation}
where $\rho$ is defined by
\[
\rho(t) = \left\{
\begin{array}{ll}
T/|t| & \mbox{for}~|t|\le T
\\
1 & \mbox{otherwise}
\end{array}
\right..
\]
Using this expression in the above integral, a straightforward scaling
argument leads to
\begin{equation}
M(2p,l) = K_{2p} \left(\frac{l}{T}\right)^{p-2p(p-1)\lambda^2} \; ,
\end{equation}
where we have denoted the prefactor
\begin{equation} 
\label{KK}
K_{2p} = T^p \sigma^{2p} (2p-1)!! \int_{0}^1 du_1...\int_{0}^1 du_p
\prod_{i<j} |u_i-u_j|^{-4\lambda^2} \; .
\end{equation}
Note that $K_{2p}$ is nothing
but the moment of order $2p$ of the random variable
$X(T)$ or equivalently of $\delta_T X(t)$.
From the above
expression, we thus obtain
\[
\zeta_{2p} = p-2p(p-1)\lambda^2
\]
and by analytical continuation, the corresponding
full $\zeta_q$ spectrum is thus
the parabola
\begin{equation}
 \label{zetamodel}
\zeta_q = (q-q(q-2)\lambda^2)/2 \; .
\end{equation}
Let us remark that one can show
that $K_{q}=+\infty$ if $\zeta_q < 0$ (i.e., $q>2+1/\lambda^2$)
and thus the pdf of $\delta_l X(t)$ have fat tails \cite{prl}. In order to
control the order of
the first divergent moment (without changing $\lambda$), one could simply
choose for
the $\epsilon_{\Delta t}$'s a law with fat tails. Indeed, the prefactor
$\sigma^{2p} (2p-1)!! $ in Eq. (\ref{KK}) comes directly from the fact
that the
$\epsilon_{\Delta t}$'s have been chosen to be Gaussian. Using instead fat
tail laws (e.g., t-student laws) would
allow us to control the divergence of this prefactor.

\begin{figure}
  \begin{center}
    \includegraphics{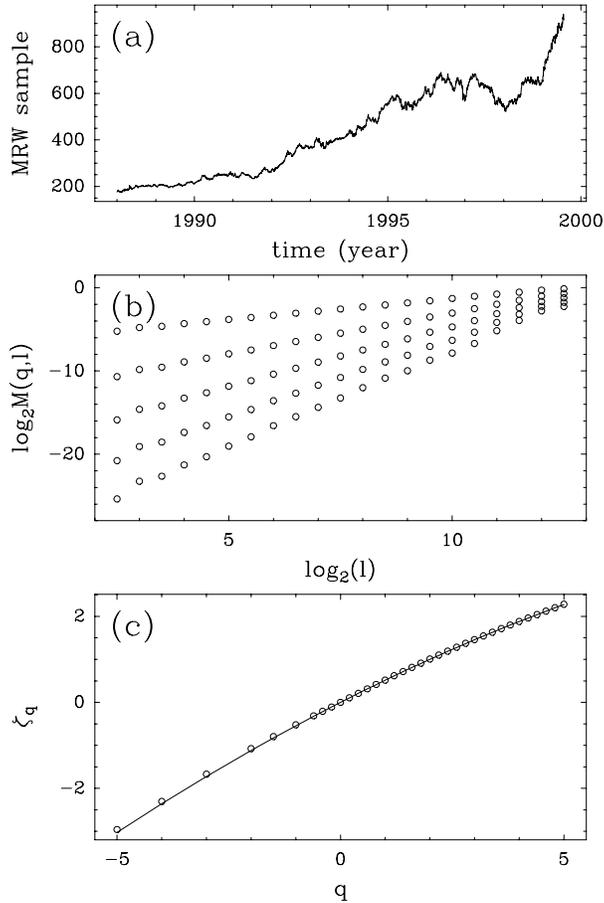}
    \caption{
      {\bf Multifractal Analysis of a MRW sample}.
      (a) Plot of a sample time serie of length $2^{17}$. The sampling
       size and the trend amplitude
       have been chosen arbitrarily to be compared to fig \ref{fig_1}(a).
      (b) Log-log plots of $M(q,l)$ versus $l$ for $q=1,2,3,4,5$.
       The time scales $l$ range from few minutes to one year.
      (c) $\zeta_q$ spectrum estimation (dots) and comparison to
      prediction as given by Eq.~(\ref{zetamodel}) (solid line).
      }  
    \label{fig_3}
  \end{center}
\end{figure}

In Fig. \ref{fig_3} we have estimated
the scaling behavior of the absolute moments $M(q,l)$ for
a discrete simulation of a MRW (Fig. \ref{fig_3}(a)).
In order to simulate the sampling of a time continuous MRW, we have
generated a discretized MRW
using $\Delta t << 1$ and then subsampled it at the sample period 1. Using
this procedure, we
have generated a $2^{17}$ long time-series using
the parameters
$\Delta t = 1/16$, $T = 2^{15}$, $\sigma^2 = 1$ and $\lambda^2 = 0.03$.
In Fig. \ref{fig_2}(b), we have plotted,
in double logarithmic representation $M(q,l)$ versus $l$
for different values of $q$. In these representations, the linear
behavior of each moment indicates that the scaling hypothesis
is verified. The estimation of $\zeta_q$ (made by estimating the slope
of each of such curve)
is reported in Fig \ref{fig_2}(c). As expected this spectrum
is a parabola that is in very good agreement with expression
(\ref{zetamodel}).

It is clear that the same power law scaling does not stand when $l$ goes
to $+\infty$. Since $\rho(l)= 1$ for large $l$ (as compared to $T$),
we get
\begin{eqnarray}
\nonumber
  E(X(l)^{2p}) &\sim_{l >> T} &
\frac{\sigma^{2p}(2p)!}{2^p p!}
\int_{0}^{l}du_1
...\int_{0}^{l}du_p \\
\nonumber
& \sim  & C l^p
\end{eqnarray}
Thus, there is a cross-over from the parabolic multifractal behavior at
time scales $l \leq T$ which is described by Eq. (\ref{zetamodel})
to the Brownian-like behavior at larger time scales ($l >> T$)
$$
\zeta_q = q/2 \; .
$$

\begin{figure}
  \begin{center}
    \includegraphics{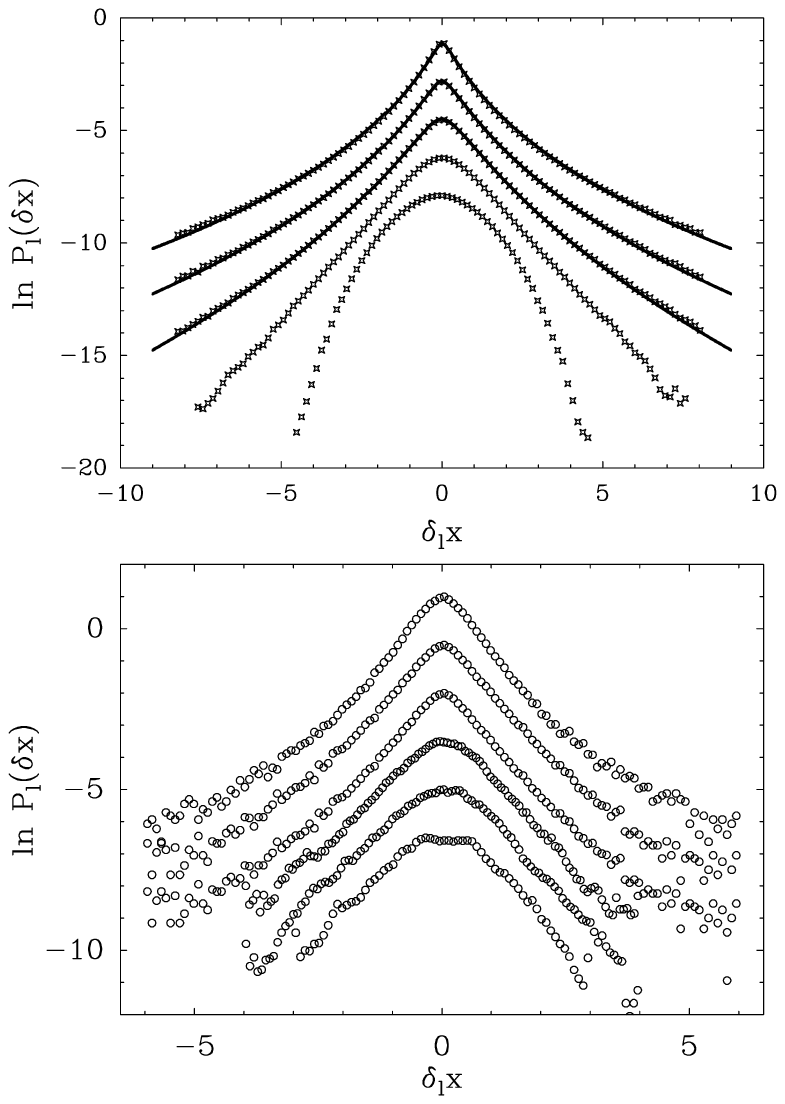}
    \caption{
      {\bf Continuous deformation of increment pdf's across scales}.
      (a) MRW Model. Standardized pdf's (in logarithmic scale)
      of $\delta_l X(t)$ for 5 different time scales (from top to bottom),
      $l=16,128,2048,8192,32768$.
      These pdf's have been
      estimated on 500 MRW realizations of $2^{17}$ sampled points
      with $\lambda^2 = 0.03$, $\Delta t = 1/16$ and
      $T = 8192$. One can see the continuous deformation
      and the appearance of fat tails when going from large to fine scales.
      In solid line, we have superimposed
      the deformation of the large scale pdf using Castaing's
      equation (\ref{castaing}) with the normal self-similarity kernel.
      This provides an excellent fit of the data.
      (b) S\&P 500 future. Standardized pdf's at scales (from top to bottom)
      $l=10,40,160$ min, 1 day, 1 week and one month.
      As in Fig. (a) the scale is logarithmic and plots have been
      arbitrarily shifted along vertical axis for illustration purpose.
      Notwithstanding the small size of the statistical sample (as compared
      to (a)), one clearly sees the same phenomenon as for the MRW.
      The fact the Castaing's equation (\ref{castaing}) allows one to
      describe the pdf's deformation across time scales of financial assets
      has originally been reported in Ref.~\cite{aGha96} where similar plots
      for FX rates can be found.
      }  
    \label{fig_4}
  \end{center}
\end{figure}

In Eq.~(\ref{kernelscaling}), we have shown that there exists a
deep link between the self-similarity kernel and the $\zeta_q$
spectrum. This suggests that
the probability distribution functions of our model
satisfy Castaing's equation when going from large to small time
scales with a gaussian kernel $G_{l,T}$. Thus, as far as the
increment pdf at different time scales are concerned,
they will satisfy an evolution equation from ``quasi-Gaussian''
at very large scale ($ l >> T$) to fat tailed pdf's at small scales.
This transformation of the pdf's is illustrated in Fig. \ref{fig_4}(a)
where are plotted, in logarithmic scale,
the standardized pdf's (the variance has been
set to one) for different time scales in the range $[1,4T]$.
The pdf's have been estimated for 500 realizations of
size $2^{17}$ of MRW with parameters $\lambda^2=0.03$ and $T=2^{13}$.
In solid line, we have superimposed the Castaing's transformation
obtained from the coarse scale pdf (at scale $T$)
using the appropriate normal self-similarity kernel.
If Fig. \ref{fig_4}(b) we have reproduced similar analysis
for the S\&P500 future variations. Besides statistical convergence
limitations,
one can observe the same features as in Fig. \ref{fig_4}(a).

\subsection{Volatility and magnitude correlation functions}

\subsubsection{Volatility correlation functions}
As recalled in the introduction, increments of financial time
series are well known to be uncorrelated
(for time lags large enough) while
their amplitude (``local volatilities'')
possesses power-law correlations.
Let us show that our model
satisfies these two properties at all time scales smaller than
the ``integral time'' $T$. By construction,
the increment correlation function,
$$\<(X_{\Delta t}(t+\tau)-X_{\Delta t}(t))( X_{\Delta
t}(t_1+\tau_1)-X_{\Delta t}(t_1))\>$$
($\forall \; |t_1-t| > \tau$),
is zero in our model. Let us
study the correlation function of the squared increments. Since the
increments are stationary, we can choose arbitrarily
$t_1 = 0$. Thus we need to compute, in the limit $\Delta t \rightarrow 0$,
the following correlation function, that corresponds to
a lag $l$ between increments of size $\tau$
\begin{equation}
\label{vcf}
C(l,\tau) =  \<(X_{\Delta t}(l+\tau)-X_{\Delta t}(l))^2 X_{\Delta
t}(\tau)^2\>.
\end{equation}
From the results of Ref. \cite{prl} and in the case
$0 \le l<T$, $0\le \tau+l<T$,
we get, 
\begin{eqnarray}
\nonumber
C(l,\tau) = \sigma^{4} \int_l^{l+\tau} du
\int_0^{\tau} dv \rho(u-v)^{4 \lambda^2}.
\end{eqnarray}
A direct computation shows that
\begin{eqnarray*}
& \int_{l}^{l+\tau} du \int_0^{\tau} dv |u-v]^{-4\lambda^2} = \\
& \frac{1}{(1-4\lambda^2)(2-4\lambda^2)} ((l+\tau)^{2-4\lambda^2} +
(l-\tau)^{2-4\lambda^2} - 2l^{2-4\lambda^2}),
\end{eqnarray*}
and consequently
\begin{equation}
C(l,\tau)  = K ( |l+\tau|^{2-4 \lambda^2} +|l-\tau|^{2-4\lambda^2}
-2 |l|^{2-4 \lambda^2} )
\end{equation}
where
$$
K =
\frac{\sigma^4 T^{4\lambda^2}}{(1-4\lambda^2)(2-4\lambda^2)}.
$$
Let us note that in the usual case $0 \le \tau << l$, one gets
\begin{equation}
C(l,\tau)  \simeq 
\sigma^4\tau^2 \left(\frac{l}{T}\right)^{-4\lambda^2}
\end{equation}
i.e., for fixed $\tau$,
the volatility correlation function scales as
\begin{equation}
  \label{c2}
   C(l) \sim  l^{-2 \nu}
\end{equation}
with $\nu = 2\lambda^2$.
From the estimates $\lambda^2 \simeq 0.025-0.05$
for financial assets (see section 5),
one thus obtains $\nu \simeq 0.05-0.1$, values very close
to the ones observed empirically in many works.

\subsubsection{Power of returns and magnitude correlation functions}
Let us now show that magnitude correlation functions behave as expected,
i.e, decrease very slowly as a logarithmic behavior.

For that purpose, the previous computation of the correlation function
can be extended to the power of returns
$|X_{\Delta t}(l+\tau) - X_{\Delta t}(\tau)|^p$. Several empirical works
have concerned the study of such ``generalized volatilities''
and people often noticed
variations of amplitude of the correlation and of
the power-law exponent $\nu_p$ when varying the order $p$ \cite{aDing96}.
In Ref. \cite{prl}, it is shown that the quantity,
\begin{equation}
C_p(l,\tau) =  \<|X_{\Delta t}(l+\tau)-X_{\Delta t}(l)|^p |X_{\Delta
t}(\tau)|^p\> \; ,
\end{equation}
behaves, when $\tau$ is small enough, as
\begin{equation}
  C_p(l,\tau) \sim K_{p}^2 \left(\frac{\tau}{T}\right)^{2\zeta_{p}}
\left(\frac{l}{T}\right)^{-\lambda^2 p^2}
\end{equation}
where the constant $K_p$ has been defined previously.
Using analytical continuation of the behavior of $C_p$ in the
limit $p = \epsilon \rightarrow 0$, we can obtain, from previous expression,
the behavior of the magnitude correlation function $C_{\omega}(l,\tau)$:
\begin{equation}
  \label{cfmag}
 C_{\omega}(l,\tau) \simeq
 \epsilon^{-2} \left(
   C_\epsilon(l,\tau) - M(\epsilon,\tau)^2 \right)
 \sim -\lambda^2 \ln(\frac{l}{T})  \; .
\end{equation}
The magnitude correlation function, for $\tau$ small enough,
has thus the same behavior as
the correlation function of the underlying
magnitude process $\omega_{\Delta t}$.
This result is checked in Fig. \ref{fig_4} where we have
plotted the magnitude correlation function for $\tau = 32 \Delta t$
as a function of $\ln(l)$. This correlation function
has been estimated using a single realization of the
process of $2^{17}$ sampled points, i.e, $16$ integral scales.
The linear behavior we obtain is exactly the same one
as predicted from Eq.~({\ref{cfmag}) and Fig.~\ref{fig_2}.
  Measures of the
slope and the intercept of such straight line provide
a good estimate of respectively
$\lambda^2$ and $T$.
\begin{figure}
  \begin{center}
    \includegraphics{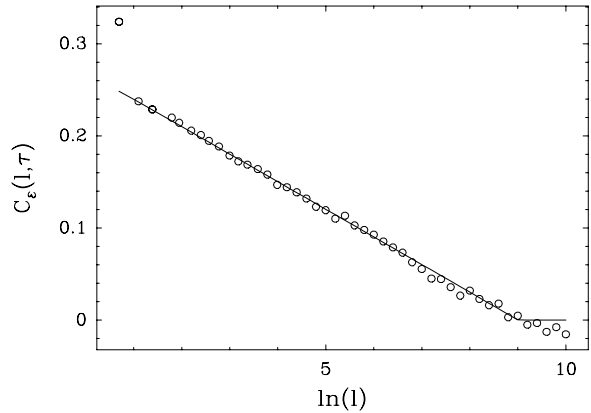}
    \caption{
      {\bf Magnitude correlation function of the model}.
      The correlation function has been estimated on 16 integral
      scales (see text). In continuous line we have superimposed
      the correlation function of the magnitude process $\omega_{\Delta t}$
      that is involved in the stochastic volatility.
     }  
    \label{fig_5}
  \end{center}
\end{figure}

\subsection{Extension to a multivariate process}
In order to account for the fluctuations of financial portfolios and
to consider management applications of our approach, it is important
to build a multivariate, i.e. a vector valued, version of the previous
multifractal random walk. Since only gaussian random variables are
involved in the construction of section 4,
this generalization can be done by considering two uncorrelated
gaussian random vectors $\ve_{\Delta t}(t)$ and
$\vo_{\Delta t}(t)$ whose covariance matrices are denoted
respectively $\mS$ and $\mL$.
Hereafter, we will refer to these matrices as respectively
the ``Markowitz matrix'' $\mS$ and the ``multifractal matrix'' $\mL$.
One can then define the multivariate multifractal random walk
(MMRW) $\vec{X}(t)$ as:
\begin{equation}
     X_{i}(t+\Delta t)-X_{i}(t) = \epsilon_{i}(t) e^{\omega_{i}(t)} \; ,
\end{equation}
with $\Cov(\epsilon_{i}(t),\epsilon_{j}(t+\tau)) = \delta(\tau) \mS_{ij}$
and $\Cov(\omega_{i}(t),\omega_{j}(t+\tau))
= \mL_{ij} \ln(T_{ij}/|\Delta t + |\tau|)$
(note that the previously defined coefficients $\sigma^2$ and
$\lambda^2$ for an asset $i$ correspond respectively to
the diagonal elements $\mS_{ii}$ and $\mL_{ii}$).
Let us briefly review some of the properties
of this model, postponing its detailed analysis to a forthcoming
publication \cite{inprep}.
A quantity that will be of central interest is the $k$-point
joint moment of order $q_1,q_2,...,q_k$ that can be defined as:
\begin{equation}
\label{jmg}
  M_{i_1,...,i_k}(q_1,...,q_k) = E\left(|X_{i_1}(l)|^{q_1}...
|X_{i_{k}}(l)|^{q_k}\right) \;.
\end{equation}
When $k=2$, by denoting $i_1=i$, $i_2=j$, $q_1=p$ and $q_2=q$,
let us define the joint scaling exponent spectrum as:
\begin{equation}
  \label{jm}
  M_{i,j}(p,q) = C_{i,j}(p,q) l^{\zeta_{i,j}(p,q)} \; .
\end{equation}
This spectrum can be computed analytically.
If the matrix $\mS$ is diagonal
(the $\epsilon_i$'s are uncorrelated), a straightforward calculation shows
that the scaling exponent
$\zeta_{i,j}(p,q)$ is the following:
\begin{equation}
\label{zetaij1}
   \zeta_{ij}(p,q) = \zeta_{i}(p)+\zeta_{j}(q) -\mL_{ij}pq \; ,
\end{equation}
where $\zeta_{i}(q)$ is the $\zeta_q$ spectrum for the component
$X_{i}(t)$. Thus, for uncorrelated $\omega_i$'s, one
has $\zeta_{ij}(p,q)=\zeta_i(p)+\zeta_j(q)$ while for
the extreme case $\omega_i=\omega_j$, the exponent becomes
$\zeta_{ij}(p,q)=\zeta_i(p+q)=\zeta_j(p+q)$.
The computation of the scaling exponent
is trickier for general Markowitz and
multifractal matrices. Under some mild conditions
that are necessary for the existence of a non trivial limit
$\Delta t \rightarrow 0^+$, one can show that the previous scaling
law remains valid even for non diagonal matrix $\mS$ \cite{inprep}.

In order to define a simple way to get
an estimate of the multifractal covariance coefficient $\mL_{ij}$, let us
define the moment ratio:
\begin{equation}
   R_{ij}(q,l) = \frac{E(|X_i(l)|^q|X_j(l)|^q)}{E(|X_i(l)|^q)E(|X_j(l)|^q)}
\sim
l^{\kappa_{ij}(q)}
\end{equation}
From Eq. (\ref{zetaij1}),  the value of $\kappa_{ij}(q)$
is simply
\begin{equation}
\label{kappa}
 \kappa_{ij}(q) =  -\mL_{ij}q^2 \; .
\end{equation}
Thus, the non-diagonal element
in the multifractal matrix $\mL$ corrresponds to the nonlinear
behavior of the exponent spectrum $\kappa(q)$ of the moment ratio $R$.
Along the same line as for the computation of the magnitude
auto-correlation in previous section, one can get
the correlation function of magnitudes $\omega_i(t,l)$ and $\omega_j(t,l)$
from the limit $q \rightarrow 0$ of $R_{ij}(q,l)$:
\begin{equation}
  \label{magcc}
   \Cov(\omega_i(t,l),\omega_j(t,l)) \sim -\mL_{ij}\ln(l)+C \; ,
\end{equation}
where $C$ is a constant related to $T_{ij}$ \cite{inprep}.
Thus the scale behavior of the magnitude covariance provides an estimate
of the multifractal correlation coefficient $\mL_{ij}$. This is the
generalization of the classical result in multifractal analysis
that relates the intermittency coefficient $\lambda^2 = \mL_{ii}$
to the scale behavior of the variance of the magnitude.

Let us remark that the covariance of
the variations of the assets $i$ and $j$
can be obtained by a direct calculation:
\begin{equation}
\label{covx}
   \Cov(X_{i}(l),X_{j}(l)) =
   \mS_{ij}e^{\frac{1}{2}(\mL_{ii}+\mL_{jj}-2\mL_{ij})}l
   \; .
\end{equation}
This covariance between $X_i(t)$ and $X_j(t)$ thus
depends not only on $\mS$, the ``Markowitz'' covariance matrix,
but also on the multifractal matrix $\mL$.
This expression, allows us to get an estimate of the value
of $\mS_{ij}$ once the values of $\mL$ are known.

Finally, let us mention that the idea of ``multivariate multifractality''
has been recently introduced in Ref. \cite{pf} where the
authors propose a phenomenological
generalization of Castaing's equation to the multivariate setting.
Evidences that financial assets are characterized by
non trivial multifractal matrices are also provided.
We are currently working to obtain further empirical evidences
towards such conclusions. Moreover, a precise link
between the present model and the
extended Castaing's approach of Ref. \cite{pf} is under progress.
\begin{table}
\begin{center}
 \begin{tabular}{|c||c|c|c|}
  \hline Series & Size & $\lambda^2$ & $T$ \\
  \hline 
  \hline Future S\&P500 & $7.10^4$ & 0.025 & 3 years \\
  \hline Future JY/USD  & $7.10^4$ & 0.02  & 6 months \\
  \hline Future Nikkei  & $7.10^4$ & 0.02  & 6 months \\
  \hline Future FTSE100 & $7.10^4$ & 0.02  & 1 year \\
  \hline S\&P500 index & $6.10^3$ & 0.024 & 3 years \\
  \hline French index & $6.10^3$ & 0.029 & 2 years \\
  \hline Italian index & $6.10^3$ & 0.029 & 2 years \\
  \hline Canadian index & $6.10^3$ & 0.024 & 3 years \\
  \hline German index & $6.10^3$ & 0.027 & 3 years \\
  \hline UK index     & $6.10^3$ & 0.026 & 6 years \\
  \hline hong-kong index &$6.10^3$ & 0.05 & 3 years \\
  \hline
\end{tabular}
 \caption{{\bf Multifractal paramater estimates for
     various assets}}
 \label{tab1}
\end{center}
\end{table}

\section{Parameter estimation for real financial data}
We have seen that the MRW is characterized mainly
by 3 parameters: $\sigma^2$, the white noise variance,
$T$ the integral scale and $\lambda^2$ the magnitude variance.
We have shown that this model is able to reproduce all the
main features of the future S\&P 500 time series.
Natural estimators of those parameters can be defined
from the results of previous section.
The parameter $\lambda^2$ can be obtained from the
shape of the $\zeta_q$ spectrum that is itself estimated using
the scaling behavior of the absolute moments $M(q,l)$.
This parameter can also be estimated thanks to the magnitude correlation
function $C_{\omega}(l,\tau)$ that behaves as $-\lambda^2\ln(l/T)$.
From the intercept of such correlation function as a function of $\ln(l)$,
we can define an estimator of the integral scale $T$. Finally, the
parameter $\sigma^2$ can be obtained using the classical relationship
$\Var(\delta_l X_{\Delta t}(t)) = \sigma^2 l$.
In this section, we report estimates of the multifractal
parameters $\lambda^2$ and $T$ for some financial
time series.
We do not have the ambition
to provide fine estimates of those parameters.
Our aim is rather to get an idea of realistical values of the
parameters of the model for real assets.
A precise discussion of the properties of various estimators
from a statistical point of view is out of the scope of this
paper and will be addressed in a forthcoming
publication. Note that similar empirical study has already been performed in
Ref. \cite{pf}. We have studied some high frequency future time series
that are sampled at a 10 min rate over the 7 years period
from 1991 to 1997.
We have also processed a
set of daily index values for 8 different countries over
the period from 1973 to 1997.
The results are reported in table \ref{tab1}.

We remark that the values of the multifractal parameter
$\lambda^2$ are all very close to $2.5 \; 10^{-2}$
(excepted for the hong-kong index). The integral time $T$ values
are centered around 3 years but with a large spread. Let us
notice that we get an estimate of $\ln(T)$ and thus the
error on the estimate of $T$ can be very large. We do
not report here the values of the errors and confidence
intervals for the proposed estimators that will be studied
elsewhere.

\section{Discussion about other approaches and findings}
In this section, we make some comments about related studies
that concern multifractals and finance.

 \subsection{Turbulence and finance}
The analogy between turbulence and finance has been originally proposed
by Ghashghaie {et al.} \cite{aGha96}. These authors proposed
to describe the pdf's of FX price changes at different time scales
in the same way physicists describe the pdf's of velocity variations
at different space separations in fully developed turbulence.
This approach naturally leads to the notions of cascading
process, Castaing's formula
and multifractality as described in section 2.
This work suggests that the key mechanism at the origin
of these observations, is an {\em information cascade}
according to which short-term traders are influenced by
long-term traders. This cascade is the analog of the Richardson's
kinetic energy cascade in turbulence where small eddies result from
the breakdown of larger ones and so on \cite{bFri95}.
If the observations
reported in Ref. \cite{aArn98} strongly support this point
of view, its quantitative understanding in
terms of ``microscopic'' mechanisms remains an open question.
In this section we would like to comment about
some criticisms that have been raised about
the analogy between turbulence and finance.
The first one concerns the power spectrum behavior in both situations
\cite{aMan96,pArn96,bMan00}. In turbulence, Kolmogorov
theory predicts a $k^{-5/3}$ power spectrum that is
confirmed in experimental situations.
In finance, since
price fluctuations are almost uncorrelated, they are
characterized by a $k^{-2}$ spectrum. For a general multifractal
process, the exponent $\beta$ of the power spectrum
behavior can be shown to be related \cite{bifc,wtmm2}
to the value of $\zeta_2$: $\beta =1+\zeta_2$.
Thus, from the cascading process point of view, nothing
prevents the exponent $\beta$ from being equal to the
exponent of the Brownian motion, i.e., $\beta=2$. In other words,
as examplified by the MRW,
a cascading process can have uncorrelated increments.
We could also remark, that in turbulence $\beta= 5/3$ has a
dimensional origin, i.e., it is the exponent of the
spatial spectrum of velocity fluctuations within an Eulerian description.
If one adopts a Lagrangian description and one is interested
by temporal fluctuations of a fluid particle velocity, then the
dimensional value of the power spectrum exponent is $\beta=2$.
Thus the value of this exponent is not a pertinent argument
to reject the analogy with turbulence.
Another difference that has been raised in \cite{bMan00} concerns
the behavior of the probability of return to origin $P_l(0)$
that has been shown to possess a scaling regime in finance
while its behavior is more complex for a turbulent velocity field.
First of all, let us point out that whatever the quantity
studied (probability of return or absolute moments), it is well
known that there is no observed well-defined
scaling regime in turbulence: the
classical ``log-log'' plots always display some curvature
across scales. This curvature is Reynolds number dependent
and several studies suggest that it vanishes, i.e. the
field is scale-invariant, only in the limit of infinite
Reynolds number \cite{bFri95,castaing,aArn98c}. However,
within the cascade paradigm and using Castaing's equation, the
scaling behavior of the probability of return to origin is
easy to show. Indeed, by setting $\delta X = 0$ in (\ref{castaing}),
one obtains, from the definition of $\zeta_q$ and the self-similarity
kernel:
\begin{equation}
  P_{l}(0) = P_T(0) \int G_{l,T}(u)e^{-u} du = P_T(0)
\left(\frac{l}{T}\right)^{\zeta_{-1}} \; .
\end{equation}
The exponent for the probability of return to origin is thus
simply $\zeta_{-1}$. For the log-normal stochastic
volatility model introduced in section 4, we thus get
\begin{equation}
    P_l(0) \sim l^{-\frac{1+3\lambda^2}{2}} \; .
\end{equation}
To conclude, neither the power spectrum exponent, nor the
scaling behavior of the probability of return to origin can
be used as argument against the existence of a cascading process
at the origin of the fluctuations of financial time series.

 \subsection{Subordinated processes. Multifractal time}
Subordinated processes are Markov processes in a time
variable $\mu(t)$ that is itself
an (increasing) random process \cite{feller}.
Such processes have been introduced in finance by Mandelbrot
and Taylor \cite{aMan67}
to account for the existence of Levy stable laws as the
result of a Brownian motion in some stochastic time.
Today, the idea of modelling financial
return fluctuations as a Brownian motion in a
``fractal time, ``trading time'' or ``financial time''
can be found in many approaches. In Refs. \cite{pFis97,aMan99},
the multifractal nature of these fluctuations has been modelled
by a (fractional) Brownian motion subordinated with a multifractal
stochastic measure. In this section, without any
concern for rigor, we would like to make a link between
our stochastic volatility approach and the multifractal time
approach of Mandelbrot and co-authors.
Let us first remark that if we drop the noise $\epsilon$ in
Eq. (\ref{model1})
and keep only the stochastic volatility $\sigma(t)$,
we can construct a stochastic measure $\mu(dt)$
that satisfies
$\mu(dt) = e^{\omega(t)} dt$.
Using exactly the same kind of computation as in
section 4, one can show that this measure
is stationary and its multifractal spectrum $\tau(q)$ is
\begin{equation}
  \label{taumu}
   \tau(q) = \zeta_{2q}-1 \; ,
\end{equation}
as usually defined by
\begin{equation}
 \label{deftau}
    \<\mu([0,t])^q\> \sim t^{\tau(q)+1} \; .
\end{equation}
Let us note that the existence and the construction of such a measure
that is stationary and possesses a continuous scale invariance, was
at the heart of the construction in
Refs. \cite{pFis97,aMan99} and was still an open problem.
According to these studies, one can thus construct a multifractal
process by simply considering the subordinated process $S(t) =
B(\mu([0,t]))$
where $B(t)$ is the standard Brownian motion. The $\zeta_q$ spectrum
of such process would be exactly the same as the stochastic
volatility process defined in section 4.
This is not so surprising since, formally,
a differential form for the process $S(t)$ would be
$dS = \frac{d\mu}{dt} dB(\mu(t))$. If one assumes
that a white noise that is subordinated remains a white noise, one thus
obtains $dS = e^{\omega(t)} dB(t)$ that is the equation that defines
the MRW of section 4.
The questions of well-definiteness of this construction,
its statistical properties and
the precise mathematical justification of such results,
will be addressed in a forthcoming work.

\subsection{Some remarks about Bouchaud, Potters and Meyer's model}
Besides multifractal and cascade pictures, our present approach has been
inspired by a recent paper by Bouchaud, Potters and Meyer \cite{aBou00}.
These authors
have proposed a model that is very similar to ours: the stochastic
volatility $\sigma(t)$ instead of being log-normal ($e^{\omega(t)}$)
is a normal ($\omega(t)$) random process
with long-range (power-law) correlations.
By a simple analytical computation,
they have shown that the q-order cumulants of such a process satisfy
a simple scaling behavior but the moments display
apparent multiscaling caused by a ``competition'' between
the different cumulant behavior on a finite scale range.
They thus conclude that a distinction between multifractality
and such ``apparent multifractality'' is a difficult task for
finite size time series. As far as multifractal analysis
and modelling of financial time series are concerned,
this work is very interesting
and the previous assertion is undoubtedly
difficult to infirm. However, let us remark that in order to illustrate
their purpose, Bouchaud {\em et al.} choose a ``stochastic volatility''
$\sigma(t)= e^{\omega(t)}$ instead of their ``monofractal''
model $\sigma(t) = |\omega(t)|$. The reason invoked by the
authors is that the log-normal is ``a more realistic time series as compared
with real data...without changing the feature of the above model, i.e.
the very slow decay of the volatility correlations''. They thus
claim that the scaling features of both models are the same
and thus that the multifractality observed
for the simulations of the ``log-normal'' volatility model is only
apparent as predicted by their theory for the ``normal'' volatility
model. The results reported in section 4 can be used to show
that this interpretation is not correct. Let us indeed reconsider
both results of Ref. \cite{aBou00} and section 4.
According to the ``normal''
volatility model, the moment of order
$q = 2p$ is written in terms of cumulants and behaves as \cite{aBou00}:
\begin{equation}
   M(2p,l = N \Delta t) = A_{2p,0} N^{(1-\nu)p} \; + \; ...\; + \;
A_{2,..,2} N^{p} \,
\end{equation}
where the constants $A_{q_1,..,q_k}$ depend only $q_i$, $\lambda^2$
the variance of $\omega$ and $\nu$ the exponent for the correlation
function of $\omega$: $C_{\omega}(l) \sim l^{-\nu}$.
According to this equation, if $N$ is small enough,
$M(2p,l) \sim l^{(1-\nu)p}$
while, for $N$ very large, $M(2p,l) \sim l^{p}$. The transition scale
$N^{\ast}(q=2p)$ above which the scaling exponent is
$\zeta_q = q/2$ can be estimated if we define it as the scale where the
contribution to the moment of order $q$ of the cumulant of order 4 and 2
are equal. Using the expression in
Ref. \cite{aBou00} for second and fourth cumulants, $C_2$ and $C_4$,
we can show that at scale $N^{\ast}(q=2p)$, we have
$(2p-1)!! C_2^{p} \simeq p(p-1)(2p-1)!! C_2^{p-2}C_4/6$.
From the value of $C_4$, we obtain ($q > 1$):
\begin{equation}
    N^{\ast}(q=2p) = \left( p(p-1) \nu^2 2^{2(\nu-1)} \sum_{m=1}^{+\infty}
      m^{2(\nu-1)} \right)^{\frac{1}{2\nu}}
\end{equation}
This function only depends on $q$ and
$\nu$ and is increasing as $q \rightarrow +\infty$.
Thus, the larger the $q$ value, the wider the range of scales
on which apparent multifractality exists.
However, a numerical computation of the values of $N^{\ast}$
for $\nu=0.2$ shows that the value $N^{\ast} \simeq 100$ is reached
only for the moment of order $p=15$. The greatest moment value
attained in practical situations is $q \simeq 6$,
for which $N^{\ast} < 1$~!
That means that for all moments less that $10$, the model of Bouchaud
Potters and Meyer predicts the trivial spectrum $\zeta_q = q/2$
without any cross-over phenomenon. For their numerical simulations, they
have used a log-normal model.
However, within the log-normal ansatz, the conclusions of Ref. \cite{aBou00}
are questionnable since, when
$\nu$ is small enough, this model is very close to the model introduced
in section 4. Let us indeed consider as in \cite{aBou00} that
$C_{\omega}(l) \sim \lambda^2 \left(\Gamma(\nu)
\cos(\frac{\pi\nu}{2})\right)^{\frac{1}{2}}l^{-\nu}$
with $\nu$ very small\footnote{Notice that there is no reason to
consider the same value of $\nu$ for the normal and log-normal
models. The results of this paper suggest that, in finance,
the value for the correlation exponent in the log-normal model is
very close to zero and significantly smaller than 0.2}.
By expanding this expression, we obtain
\begin{equation}
  C_{\omega}(l) = \frac{\lambda^2}{\nu}-\lambda^2\ln(l) + O(\nu\ln(l)) \; .
\end{equation}
If we set $T = e^{\nu^{-1}}$, then for $1 \leq l << T$,
this equation becomes
\begin{equation}
  C_{\omega}(l) = \lambda^2 \ln(T/l)
\end{equation}
that is the same correlation function as introduced for
the multifractal model in section 4. Let us notice that
for $\nu=0.1$ we have $T \approx 2.10^4$, $T \approx 3.10^5$ for
$\nu=0.08$ and $T \approx 5.10^8$ for $\nu=0.05$ !.
In this model $T$ is increasing very fast as $\nu$ goes to zero.
We can thus conclude, that the model numerically
studied in Ref. \cite{aBou00} can be seen as {\em multifractal}
from one point of view:
whatever that scaling range $[1,T]$, there exists
$\nu$ small enough ($\nu \approx 1/\ln(T)$) such that the model displays
multiscaling with log-normal $\zeta_q$ spectrum in this scale range.
According to these remarks, we thus think that the multifractal
picture is more realistic to describe multiscaling in financial
time series.

\section{Summary and prospects}
In this paper we have reviewed what are the
main features of multifractal processes.
We have shown that the {\em Multifractal Random Walk}
is a very attractive alternative to classical cascade
processes in the sense that it is stationary, continuously
scale-invariant and formulated using a simple
stochastic evolution equation.
As a model for financial engineering, MRW are interesting
for many reasons. First, as illustrated in
details for the S\&P 500 intraday time series, this
model is able to reproduce the main empirical properties
observed for financial time series. Moreover, as
Brownian motion and other stable walks, it is a ``scale-free''
model in the sense that it does not have to fit a particular
time-scale since it is scale-invariant. This kind
of stability with respect to time ``aggregation''
is a serious advantage as compared to classical
ARCH-like models which parameters strongly depend
on the time-scale one is interested in.
Moreover, as discussed in section 4.4, a simple
multivariate formulation of MRW can be proposed.
To our knowledge, it is the first example
of an extension of the notions of multifractality
to a vector field.
The empirical results reported in Ref. \cite{pf}
suggest that MMRW can be pertinent for portfolio theory.
We are currently working on applications of MRW
to classical problems of finance
like management problems and option pricing theory.

From a theoretical point of view, MRW can be seen
as the simplest model that contains the main
ingredients for multifractality.
In that respect, it
can be very helpful to elucidate, in many fields
where multiscaling is observed, what are the
generic mechanisms that are involved leading
to ``non-trivial'' self-similarity properties.
Various ``microscopic'' models, as proposed
in finance or other fields, could be
considered within this perspective.
It could also be interesting to recast our approach
within a field theoretical formulation involving some
renormalization procedure.
From a mathematical point of view, this problem is
deeply linked to the existence
of a limit stochastic process when the
sampling time $\Delta t$ goes to zero. The convergence
of the moments is not sufficient to prove this non trivial
assertion. Such a limit could be very useful to
develop a new stochastic calculus within which, for example, one could
formulate the model of multifractal time
of Mandelbrot and co-authors very naturally (see section 6.2).
Finally, in a forthcoming work, we will
discuss the generalization of such approach to other laws
than the (log-)normal. \\

\noindent
{\bf Acknowledgement\\}
  We acknowledge Matt Lee and Didier Sornette for the
  permission to use their financial data. We are also
  very grateful to Alain Arneodo, Jean-Philippe Bouchaud
  and Didier Sornette for interesting discussions and
  remarks.

  All the computations in this paper have been made using
  the free GNU licensed sofware {\em LastWave} \cite{lw}.

\end{document}